\documentclass[10pt, conference, doublecolumn]{IEEEtran}

\usepackage{macros}

\usepackage{fancyhdr}
\fancypagestyle{acceptancenote}{
    \fancyhf{} 
     
    \fancyhead[L]{\footnotesize \itshape To appear in IEEE Globecom, 2026.} 
}

\title{HRRC on the Farm: Quantile Forecasting for Highly-Reliable Remote Control via LEO Networks}
\author{André Gomes}
\date{October 2025}

\begin{document}

\author{
  \IEEEauthorblockN{
    André Gomes\IEEEauthorrefmark{1}, 
    Jie Wang\IEEEauthorrefmark{2}
  }
  \\\IEEEauthorblockA{
    \IEEEauthorrefmark{1}\textit{Rowan University}, USA, E-mail: gomesa@rowan.edu \\
    \IEEEauthorrefmark{2}\textit{Iowa State University}, USA, E-mail: jiew@iastate.edu \\ 
    % \IEEEauthorrefmark{3}\textit{AT\&T Labs}, USA%, E-mail: 
  }
}

\maketitle

\thispagestyle{acceptancenote}

% Abstract
\begin{abstract}
LEO satellite networks are an attractive solution to support farm automation in Agriculture 4.0 because of their ubiquitous coverage. 
However, LEO networks often suffer from high latency volatility, which can limit their utility in mission-critical farm operations such as remote control. 
This paper studies highly-reliable remote control over LEO networks by \1 casting highly-reliable remote control as a quantile forecasting problem and \2 proposing a high-quantile estimator that can predict latency spikes at a given reliability level. 
Our results, drawn from a real-world OneWeb dataset collected in a major agricultural hub in the US, show that the proposed estimator can support highly-reliable remote control on the farm by meeting reliability requirements while allowing the remote-controlled vehicle to operate at speeds up to $138.6\%$ higher than what would be possible otherwise. 
\end{abstract}
\begin{IEEEkeywords}
Reliability, quantile forecasting, smart agriculture, {LEO} satellite networks
\end{IEEEkeywords}
\IEEEpeerreviewmaketitle

\section{Introduction}

% \Ac{HRRC} 
% \todo[inline]{HRRC definition should appear much earlier... preferably in the first sentences of first paragraph.}

Farm automation is a key pillar of Agriculture 4.0, which includes the ability to remote control farming machinery to enhance productivity and efficiency \cite{santos2020agriculture}. 
Remote control often relies on connectivity to enable real-time data exchange, cloud-based analytics, and intelligent decision-making and operation. 
Terrestrial networks such as 5G can provide good coverage over urban areas, but they often suffer from coverage gaps in rural regions \cite{yaacoub2020key}, making them unsuitable for mission-critical farm operations such as remote control. 
\Ac{LEO} satellite networks, such as Starlink and OneWeb, have emerged as a complementary choice by providing ubiquitous coverage without the need for expensive terrestrial tower infrastructure \cite{lopez2023connecting}. 

Reliability is another fundamental requirement for ensuring the operational integrity of remote-controlled systems, where an unexpected delay in command delivery can cause a vehicle to operate under unsafe conditions (e.g., move at risky speeds or in the wrong directions), potentially resulting in crop damage, equipment collisions, or even injury to nearby animals or humans. 
Therefore, a desirable feature in mission-critical farm applications such as remote control is the ability to predict latency spikes with high confidence to ensure \ac{HRRC}, where ``highly'' refers to reliability levels that approach $1$ (or $100\%$). 
This, however, can be challenging to achieve in \ac{LEO} networks due to their inherently high mobility and dynamic nature, with a number of studies reporting on their high latency volatility \cite{husseyn2025characterizing, 10228912, pan2024measuring, mohan2024multifaceted, starlink-quantile-analysis, dataset}. 

In this paper, we consider quantile forecasting as the means of achieving safe and efficient \ac{HRRC} over \ac{LEO} networks by predicting high quantiles of future latency (e.g., $90$th or $99$th percentiles). 
Prior studies have investigated ways of coping with high latency and throughput volatility in \ac{LEO} networks. 
For instance, based on the observation that inter-satellite handovers are a major cause of high volatility, SaTCP proposed in \cite{cao2023satcp} inhibits congestion window reduction during handovers, avoiding undesirable and unnecessary throughput slowdowns.
Targeting low-latency live video streaming over Starlink networks, the algorithm in \cite{zhao2024low} utilizes contextual multi-armed bandits to adapt stream bitrate in the context of fluctuating latency and regular satellite handovers. For general latency and throughput prediction, ASTM in \cite{Tian2025rttprediction} and StarNet in \cite{liu2025vivisecting} are both deep learning-based approaches that incorporate satellite motion patterns to learn nonlinear temporal dependencies in latency and throughput, respectively, for dynamic predictions over time.
% For instance, based on the observation that inter-satellite handovers are a major cause of high volatility, \cite{cao2023satcp} proposed an extension to TCP's CUBIC that inhibits congestion window reduction during handovers, avoiding undesirable and unnecessary throughput slowdowns; \cite{zhao2024low} used machine learning to improve quality-of-experience in low-latency live video streaming over Starlink by adapting stream bitrate, while \cite{liu2025vivisecting} proposed a machine learning estimator to dynamically predict Starlink's expected throughput over time. 
While very relevant, these works focus on improving or predicting overall performance, as opposed to tail performance, a problem that has been underexplored but is highly relevant to mission-critical applications such as remote control on the farm. 

\begin{figure}[t]
    \centering
    \includegraphics[width=0.9\linewidth]{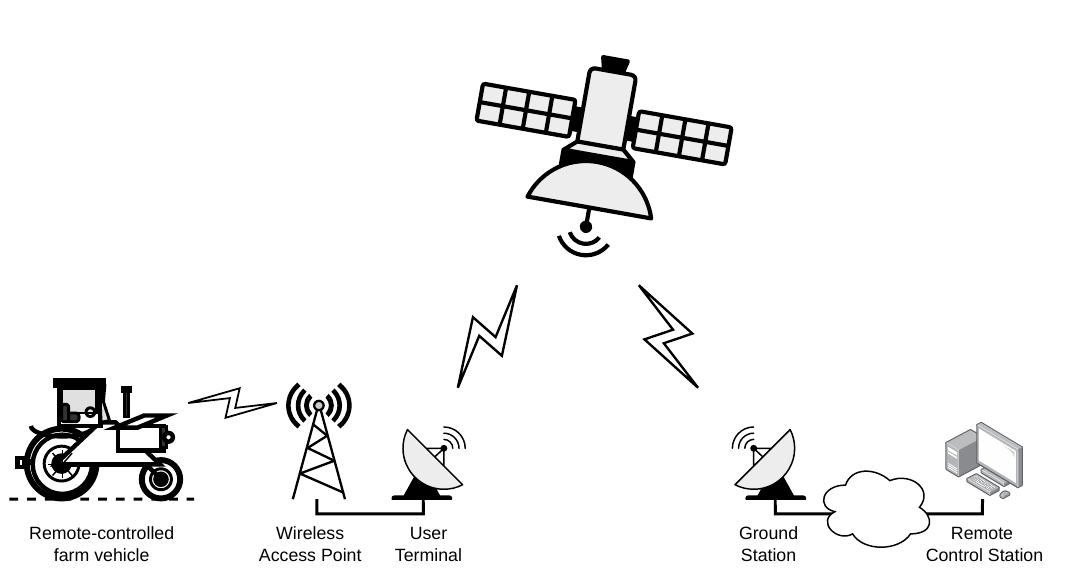}
    \caption{Farm vehicle with remote control over \ac{LEO} networks.}
    \label{fig:overview}
    \vspace{-1em}
\end{figure}

The only quantile estimator for LEO networks, to the best of our knowledge, was recently proposed in \cite{starlink-quantile-analysis},
% On that note, \cite{starlink-quantile-analysis} proposed a quantile estimator to 
which focuses on Starlink networks and predicts high quantiles through statistical characterization of latency. 
The estimator is based on the observation that Starlink handovers are often associated with distributional shifts in latency and that inter-satellite handovers in Starlink occur at well-defined intervals (every $15$ seconds) \cite{husseyn2025characterizing, 10228912, pan2024measuring, mohan2024multifaceted}. 
Soon after a handover interval times out, the proposed estimator starts collecting latency measures, which are fit into a pre-defined distribution (e.g., Gaussian or Generalized Pareto). The fitted model is used to estimate tail quantiles in between handover events. 
Useful as it is, this approach has major limitations to remote control. 
First, as per their analysis, it can take up to several seconds to collect enough measurements to fit models, a period during which remote control may not be possible. %, a limiting factor when handovers also occur in a time scale of seconds. 
Second, their approach focuses on stable intervals in between handovers, ignoring performance during handover events, which are arguably more critical for remote control. 
Third, it relies on the assumption of well defined handover intervals, which has been repeatedly observed in Starlink but not in other \ac{LEO} constellations such as OneWeb \cite{dataset}. 

In this paper, we propose a high-quantile estimator for quantile forecasting for mission-critical applications that addresses the above shortcomings. 
Specifically, our estimator leverages neural networks to predict the tail statistics of the \ac{RTT} over OneWeb, without making any assumptions on handover intervals or underlying distributions of the \ac{RTT}.  
Our estimator can predict future \ac{RTT} quantiles during and in-between handovers without incurring remote-control downtime. 
To assess its performance, we use a real-world OneWeb dataset collected in Ames, Iowa, a major agricultural hub in the US. 
Our results show that the proposed estimator can support \ac{HRRC} on the farm by meeting high-reliability requirements while allowing remote-controlled vehicles to operate at speeds up to $138.6\%$ faster than what would be possible otherwise. Our main technical contributions can be summarized as follows:
\begin{itemize}
    \item The first to model and study \ac{HRRC} for farm automation over \ac{LEO} networks. 
    \item A new \ac{RTT} quantile estimator tailored for mission-critical applications over OneWeb.
    \item Real-world evaluation based on OneWeb traces collected in a major agricultural hub in the US. 
\end{itemize}

The remainder of this paper is organized as follows. 
\Sec{system-model} presents the system model and the problem formulation to study \ac{HRRC} over \ac{LEO} networks. 
\Sec{proposed} discusses the proposed estimator for quantile forecasting as well as a new training technique for high-quantiles. 
\Sec{dataset} describes the real-world OneWeb dataset used for evaluation in \Sec{evaluation}. 
The paper concludes with a summary of main findings.

\section{System model} \label{sec:system-model}

\begin{figure}[t]
    \centering
    \includegraphics[width=0.8\linewidth]{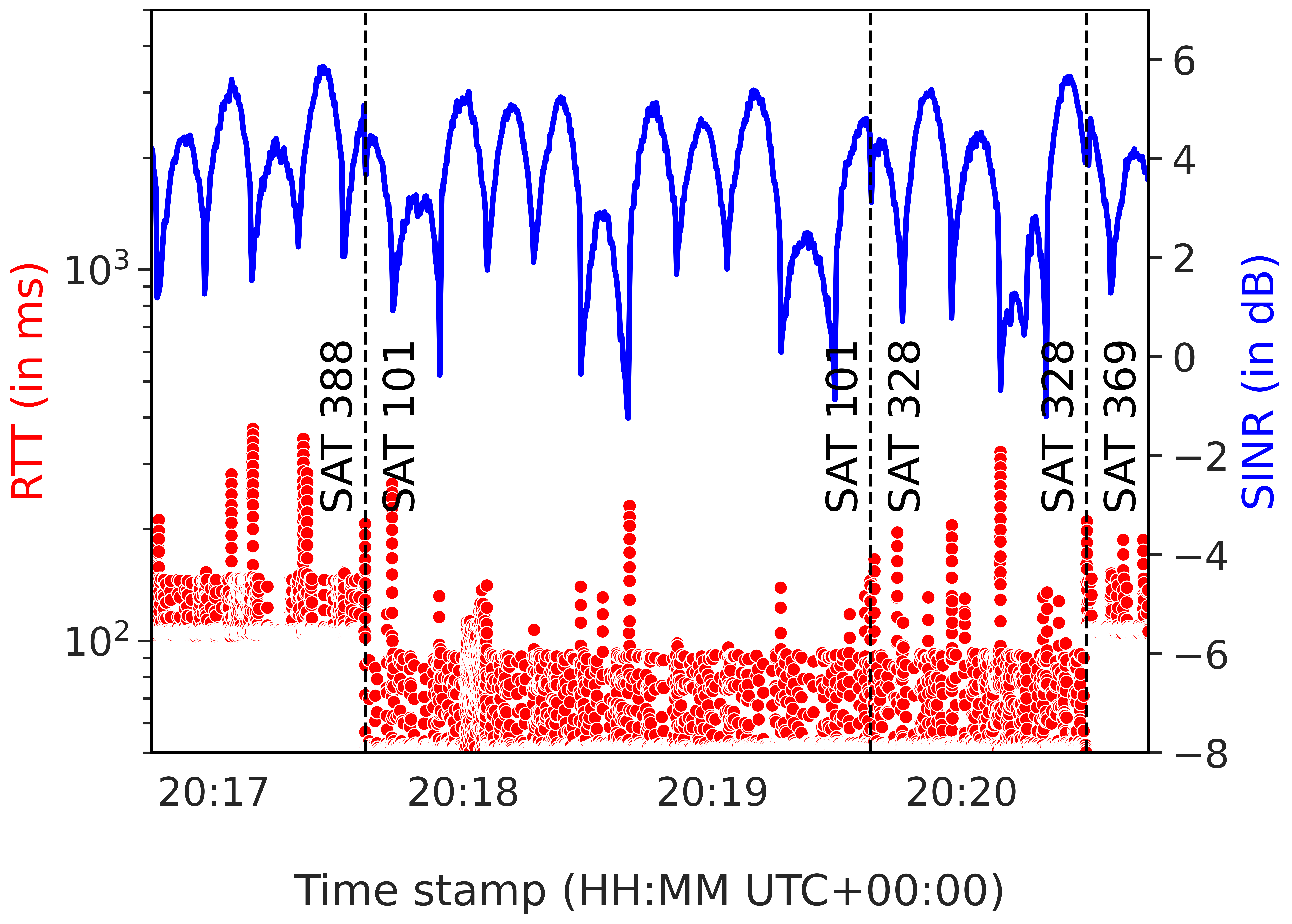}
    \caption{Snapshot of \acs{RTT} and \acs{SINR} traces from our dataset.}
    \label{fig:example}
    \vspace{-1em}
\end{figure}

We consider the network illustrated in \Fig{overview}. 
The remote-controlled farm vehicle is controlled through a \ac{LEO} network from a remote control station. 
The vehicle periodically sends its location to the control station, which responds with a control message containing, for example, updates on the vehicle's trajectory and/or speed. 
Due to network latency, the control station has to account for the vehicle's position uncertainty. 
At time $t$, the distance error $D_t$ is defined as the difference in location between the time the location is sent and the time the control message is received, and expressed as the product of the vehicle's speed and \ac{RTT}\footnote{Notation remark: $\text{RTT}_t$ and $D_t$ are random processes indexed by $t$.}:
\begin{equation}
    D_t = v_t\times \text{RTT}_t.
\end{equation}
The \emph{reliability} level $\alpha$ is defined as the probability that the distance error will not exceed a safety threshold $\epsilon$: 
\begin{equation}
    \Pr(D_t \le \epsilon) \ge \alpha.
    \label{eq:reliability}
\end{equation}
For safe operation, high-reliability levels $\alpha \rightarrow 1$ are desirable, noting that the exact reliability level depends on the scenario, from lower levels (e.g. $\alpha = 0.9$) when the vehicle is operating in the open field to higher levels (e.g., $\alpha = 0.99$) when humans, animals, or other machinery are present. 

% \subsection{Problem formulation} \label{problem}

Assuming no control over the \ac{LEO} network, the control station ensures safe and efficient operation by limiting the vehicle's speed to $v^\star_t$, the maximum speed that satisfies \Eqs{reliability}, an objective that can be formally expressed as follows:
\begin{equation}
    \begin{split}
        v^\star_t = &\max_v\left\{ v: \Pr(D_t \le \epsilon) \ge \alpha \right\}\\
        = & \max_v\left\{ v: \Pr\left(\text{RTT}_t \le \frac{\epsilon}{v}\right) \ge \alpha \right\}.
    \end{split}
    \label{eq:opt-speed}
\end{equation}
By parameterizing $z = \epsilon/v$ and thus $z^\star = \epsilon/v^\star$, this objective can be re-written in terms of the $\alpha$-quantile of the \ac{RTT}: 
\begin{equation}
    \begin{aligned}
        z^\star_t &= \min_z \left\{ z : \Pr\left( \text{RTT}_t \le z \right) \ge \alpha \right\} = F^{-1}_{\text{RTT}_t}(\alpha),
    \end{aligned}
    \label{eq:opt1}
\end{equation}
where $F^{-1}(.)$ denotes the inverse \ac{CDF}, or quantile function. 
% In other words, we are primarily concerned with upper \ac{RTT} quantiles in remote control rather than mean \ac{RTT}. 

If the distribution of the \ac{RTT} is stationary, the $\alpha$-quantile can be easily approximated numerically based on previously collected \ac{RTT} data samples (e.g., using any of the numerical approaches in \cite{hyndman1996sample}). 
In practice, however, the distribution of the \ac{RTT} is seldom stationary because of the high mobility nature of \ac{LEO} networks. 
To illustrate this, consider \Fig{example}, which shows a snapshot of \ac{RTT} and \ac{SINR} traces observed by an user terminal connected to the OneWeb constellation (see \Sec{dataset} for details). 
By visual inspection, we can observe two types of distributional shifts. 
First, note how the bulk of \ac{RTT} values shift from approximately $100~\si{\ms}$ to approximately $50~\si{\ms}$ after the inter-satellite handover between times 20:17 and 20:18; and how it shifts again after the inter-satellite handover between 20:20 and 20:21. 
Second, we can observe a concentration of higher \ac{RTT} values during \ac{SINR} dips, which are often associated with inter-beam handovers in OneWeb \cite{dataset}. 
These indicate that the distribution of the \ac{RTT} can change with handover events. 
Importantly, we note that inter-beam and inter-satellite handovers can occur frequently, on time scales of seconds to a few minutes in OneWeb \cite{dataset}, making it challenging to track \ac{RTT} quantiles over time.

\subsection{Problem statement} \label{sec:problem-definition}

To account for distributional shifts, we re-formulate the objective in \Eqs{opt1} as a \emph{conditional} quantile forecasting problem. 
Specifically, let $\mathbf{x}_t$ denote a feature vector of network measures (e.g., \ac{SINR} and \ac{RTT} measures) obtained over the past $\delta$ seconds. 
At time $t$, our goal is to predict the $\alpha$-quantile of the \ac{RTT} for a \emph{horizon} window $[t, t+\Delta)$ given network measures obtained during a \emph{historical} window $[t-\delta, t)$. 
Formally, we can express this problem as finding the conditional quantile function $F^{-1}_{\text{RTT}}(\alpha|\mathbf{x}_t)$. 

By re-writing \Eqs{opt-speed} and \Eqs{opt1} in terms of the conditional quantile of the \ac{RTT}, we define $v_t^\star$ as the maximum operational speed during the horizon window $[t, t+\Delta)$:
\begin{equation}
    v^\star_t = \frac{\epsilon}{z_t^\star} = \frac{\epsilon}{F^{-1}_{\text{RTT}}(\alpha|\mathbf{x}_t)}. 
    \label{eq:conditional-quantile}
\end{equation}
% \vspace{0.1em}

\section{Proposed quantile estimator for \acs{HRRC}} \label{sec:proposed}

We propose a neural network-based high-quantile estimator to address the problem in \Eqs{conditional-quantile}. 
Our goal is to build a model $f(\mathbf{x}_t, \alpha; \theta)$, parametrized by $\theta$, that approximates the conditional quantile function $F_{\text{RTT}}^{-1}(\alpha|\mathbf{x}_t)$. 
Note that, unlike classic forecasting problems, where a mean estimator can be obtained by training a neural network to minimize the mean squared error loss over a training set with samples of the expected output, quantile forecasting is challenging because true conditional quantiles are unknown. 

\begin{figure}[t]
    \centering
    \includegraphics[width=1\linewidth]{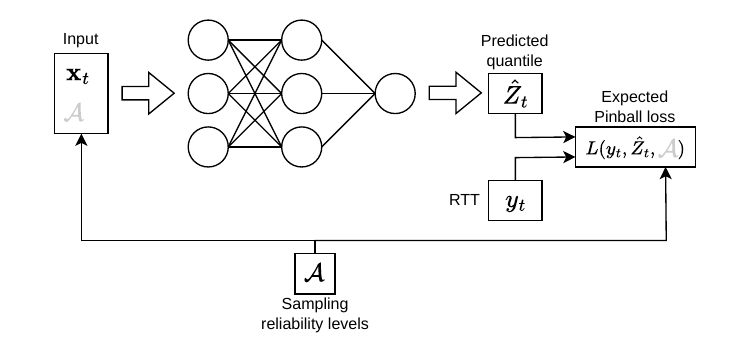}
    \caption{Example of a neural network being trained with the expected pinball loss function for quantile forecasting.}
    \label{fig:expected_pinball_arch}
    \vspace{-1em}
\end{figure}

Inspired by \cite{tagasovska2019single, expected-pinball}, we approximate the conditional quantile function by training a neural network to minimize the \emph{expected pinball loss} function over a training set $\{(\mathbf{x}_t, y_t)\}_t^T$:
\begin{equation}
    L(y_{t}, \hat{Z}_t, \mathcal{A}) =
    \begin{cases}
         \mathcal{A}\times (y_{t} - \hat{Z}_t),~\text{if}~y_{t} \ge \hat{Z}_t,\\ 
         (\mathcal{A} - 1) \times (y_{t} - \hat{Z}_t), ~\text{otherwise},
    \end{cases}
    \label{eq:expected-pinball}
\end{equation}
where $y_t$ corresponds to the observed \ac{RTT} and $\hat{z}_t = f(\mathbf{x}_t, \alpha; \theta)$ the estimated $\alpha$-quantile. 
To approximate the conditional quantile \emph{function}, $\alpha$ values are sampled at random for each training sample. 
Let $\mathcal{A}$ denote the random variable used to model $\alpha$ during training. 
Note that, if $\mathcal{A}$ is a random variable, so is $\hat{Z}_t = f(\mathbf{x}_t, \mathcal{A}; \theta)$. 
Thus, the model $f$ is obtained by finding parameters $\theta$ that minimize \Eqs{expected-pinball}: 
\begin{equation}
    \theta = \argmin_{\theta'}\left\{ \mathbb{E}\left[ \frac{1}{T} \sum_t^T L(y_t, \hat{Z}_t, \mathcal{A}) \right] \right\}. 
\end{equation}
An overview of the training process is illustrated in \Fig{expected_pinball_arch}. 

As proposed in \cite{tagasovska2019single}, a natural choice of distribution is $\mathcal{A}\sim\text{Uniform}[0, 1]$, which allows training $f$ over the entire range of $\alpha$ values. 
While this approach can be used to approximate the general conditional quantile function, we note that it often fails to capture its tails, such as when $\alpha \rightarrow 1$. 

\subsection{New training approach for high-quantile forecasting}

Augmenting the work in \cite{tagasovska2019single, expected-pinball}, we propose a new modeling for the distribution of $\mathcal{A}$ that allows the model $f$ to focus on high quantile levels $\alpha \rightarrow 1$. 
To that end, let $\text{Beta}(m, c)$ denote a Beta distribution parametrized by mode $m$ and concentration $c$, a measure of how the distribution concentrates around its mode. 
Mode and concentration can be mapped onto Beta shape parameters as follows: $a = m\times(c-2)$ and $b = (1-m)\times (c-2) + 1$. 

To focus on high quantiles, we propose modeling $\mathcal{A}$ as a mixture Beta distribution:
\begin{equation}
    \mathcal{A} \sim 
    \begin{cases}
        \text{Beta}(m_1, c_1),~\text{with probability $p_1$},\\
        \text{Beta}(m_2, c_2),~\text{with probability $p_2$},\\
        ... \\
        \text{Beta}(m_k, c_k),~\text{with probability $p_k$},
    \end{cases}
\end{equation}
with $\sum_i p_i = 1$. 
The mixture Beta distribution is used to \emph{anchor} $\mathcal{A}$ at reliability levels of interest during training. For instance, for a reliability region of interest $[0.9, 0.99]$, a natural choice of anchors for $\mathcal{A}$ are the $0.9$- and $0.99$-reliability levels:
\begin{equation}
    \mathcal{A} \sim 
    \begin{cases}
        \text{Beta}(m = 0.9, c_1),~\text{with probability $p$},\\
        \text{Beta}(m = 0.99, c_2),~\text{with probability $1-p$}.
    \end{cases}
    \label{eq:example-A}
\end{equation}
Concentration and $p$ values are hyper-parameters. 
Intuitively, higher quantile levels are more challenging to capture and train. 
The hyper-parameter $p$ allows us to prioritize higher quantile levels (such as $0.99$ in the above example) by making them more likely to be sampled during training.  
In \Sec{evaluation}, we show that the proposed high-quantile estimator can ensure safer and more efficient remote control compared to the baseline $\mathcal{A} \sim \text{Uniform}$, even when $\mathcal{A} \sim \text{Uniform}$ is modified to focus on high quantiles, e.g., $\mathcal{A} \sim \text{Uniform}[0.9, 0.99]$.

\section{Dataset} \label{sec:dataset}

To study remote control through \ac{LEO} networks, we consider a real-world OneWeb dataset from \cite{dataset, dataset-ref2}. 
The dataset consists of \ac{SINR} and \ac{RTT} traces collected from a user terminal installed in Ames, Iowa, and connected to the OneWeb satellite constellation. 
We consider traces collected from February 2025 through April 2025. 
\Ac{SINR}, obtained from the user terminal's API, is an important feature for measuring the link quality of the connected LEO satellite.
% were obtained from the user terminal's API, while
\ac{RTT} traces were captured by sending ping requests from the user terminal in Ames to the associated OneWeb PoP located in Ashburn, Virginia, approximately $1400~\si{\km}$ away from the user terminal. 

With respect to \Fig{overview}, we assume that the control station is located in Ashburn, a major data center hub in the US, while the user terminal is located in Ames, a major agricultural hub in the US. 
Communication between the remote-controlled vehicle and user terminal is handled by a private wireless network. 
While the \ac{RTT} is ultimately end-to-end, we note that delays between user terminal and remote-controlled vehicle can be fairly deterministic (e.g., using private 5G networks) and fairly small compared to delays in the non-terrestrial network. 
For simplicity, we assume that delays between user terminal and control station dominate the \ac{RTT}, while other sources of delay are assumed to be negligible. 
This way, we use \ac{RTT} traces from \cite{dataset, dataset-ref2} to characterize the communication system in \Fig{overview}.

\subsection{Data cleaning} 

After inspecting \ac{SINR} and \ac{RTT} datasets, we observed that they were collected at different sampling frequencies, $71~\si{\Hz}$ for \ac{RTT} versus $5~\si{\Hz}$ for \ac{SINR}. 
It is also worth mentioning that these datasets were timestamped with different timezones. 
To harmonize and combine the two datasets, we \1 converted timestamps to UTC+00:00 and \2 split data into discrete time steps of length $1~\si{\sec}$ (data aggregation is discussed in the following section). 
Furthermore, we note that \ac{RTT} samples were timestamped with respect to the time ICMP echo replies were received. 
We converted it to the time ICMP echo requests were sent by subtracting \acp{RTT} from the recorded timestamps. 
Lastly, we had to partially discard data based on the observation that both datasets contained gaps, likely caused by disruptions during measurements. 
We only considered non-empty intervals of length $\delta + \Delta$ seconds. 

\subsection{Final dataset}

The final dataset consists of samples $\{(\mathbf{x}_t, y_t)\}_t^T$. 
The length of the horizon window is set to $\Delta = 1$ \si{\sec}. 
For a time step $t$, the target value $y_t$ corresponds to the maximum \ac{RTT} observed over the interval $[t, t+\Delta)$. 
This way, the conditional $\alpha$-quantile in \Eqs{conditional-quantile} can be interpreted as the highest \ac{RTT} that may be observed during the horizon window with a probability $\alpha$. 

The length of the historical window is set to $\delta = 10$ \si{\sec}. 
For each time step $t$, the feature vector $\mathbf{x}_t$ contains \ac{SINR} and \ac{RTT} measures calculated over the interval $[t-\delta, t)$. Each $\mathbf{x}_t$ contains nine measures: 
\begin{itemize}
    \item \emph{(1)} average, \emph{(2)} minimum, and \emph{(3)} maximum \ac{SINR},
    \item \emph{(4)} first and \emph{(5)} second derivatives of the \ac{SINR},
    \item \emph{(6)} average, \emph{(7)} minimum, and \emph{(8)} maximum \ac{RTT}, and
    \item \emph{(9)} queuing delay. 
\end{itemize}
Queuing is estimated as the difference between the latest \ac{RTT} sample and the minimum \ac{RTT} sample in the historical window. 
Intuitively, this queuing measure is used to indicate potential \ac{RTT} peaks caused by other network conditions rather than \ac{SINR}-based delays. 
First and second derivatives of the \ac{SINR} are used to indicate imminent saddle points, which are often caused by handover events and accompanied by \ac{RTT} peaks in OneWeb \cite{dataset}. 
Derivatives are computed numerically using the Savitzky-Golay filter implemented in \cite{scipy_savgol_filter}. 

% \todo[inline]{Link intuition for feature selection to \Fig{overview}.}

\section{Evaluation} \label{sec:evaluation}

The dataset is split into three non-overlapping sets: \1 the training set contains traces collected in February 2025 (a total of 661,950 $\{(\mathbf{x}_t, y_t)\}$ samples); \2 the validation set, traces collected in March 2025, which are used for early stopping and hyper-parameter tuning (1,103,363 samples); and \3 the test set, traces from April 2025, which are used for evaluation only (1,244,899 samples). 
For each evaluation scenario, we \1 train 50 estimators over $300$ epochs, a number that was selected based on the observation that validation error did not decrease significantly after $300$ epochs, and \2 report on the average performance over the 50 independently trained estimators. 
Each estimator consists of a neural network with two fully connected hidden layers with 256 neurons each. Each hidden layer is coupled with a ReLU activation function. Neural networks are trained with PyTorch's Adam optimizer and a learning rate of $10^{-4}$. 
In terms of data normalization, we apply the Z-score normalization to feature vectors $\mathbf{x}_t$ and the $\log_{10}(.)$ normalization to target values $y_t$. 
% The length of historical and horizon windows are set to $\delta = 10$ \si{\sec} and $\Delta = 1$ \si{\sec}. 

We focus on the range $\alpha \in [0.9, 0.99]$ and model $\mathcal{A}$ as in \Eq{example-A}. 
Hyper-parameters were obtained empirically by assessing combinations of $c_1 \in \{10, 100, 1000\}$, $c_2 \in \{10, 100, 1000\}$, and $p \in \{0.1, 0.2, ..., 0.9\}$. 
By analyzing their performance with respect to the validation set, we found that the combination $(c_1, c_2, p) = (100, 100, 0.1)$ yielded the best performance according to the following criterion: for each combination, we performed the accuracy analysis in \Sec{accuracy}, calculated the $R^2$ score %\cite{r2-score} 
with respect to ``Ideal'', and selected the combination the yielded the lowest $R^2$ score, i.e., the combination that performed the closest to ``Ideal''.  
It is worth noting that $p = 0.1$ indicates that the distribution anchored at the $0.99$ is sampled more frequently, which is expected since higher quantiles tend to be more challenging to capture and train for. 
In the following sections, we report on the performance of the proposed estimator for $(c_1, c_2, p) = (100, 100, 0.1)$.  

To assess the performance of the proposed high-quantile estimator, we consider the following baselines:
\begin{enumerate}
    \item The \textbf{uniform} estimator refers to the vanilla approach proposed in \cite{tagasovska2019single} in which neural networks are trained with the expected pinball loss function with $\mathcal{A} \sim \text{Uniform}$. We consider two variations:
    \begin{enumerate}
        \item $\mathcal{A} \sim \text{Uniform}[0, 1]$, and
        \item $\mathcal{A} \sim \text{Uniform}[0.9, 0.99]$. 
    \end{enumerate}
    \item The \textbf{unconditional} estimator refers to the classic strategy of approximating quantiles numerically on the overall distribution. 
    Let $\mathcal{Y} = \{y_1, y_2, ..., y_n\}$ denote the entire set of \ac{RTT} samples in the training set. 
    The unconditional $\alpha$-quantile is approximated by $F^{-1}_{\text{RTT}_t}(\alpha) \approx F^{-1}_{\mathcal{Y}}(\alpha)$ %using \cite{numpy-quantile}
    , where $F^{-1}_{\mathcal{Y}}(\alpha)$ is calculated as in \cite[Definition 3]{hyndman1996sample}. 
\end{enumerate}

\subsection{Quantile forecasting}

\begin{figure}[t]
    \centering
    \includegraphics[width=0.8\linewidth]{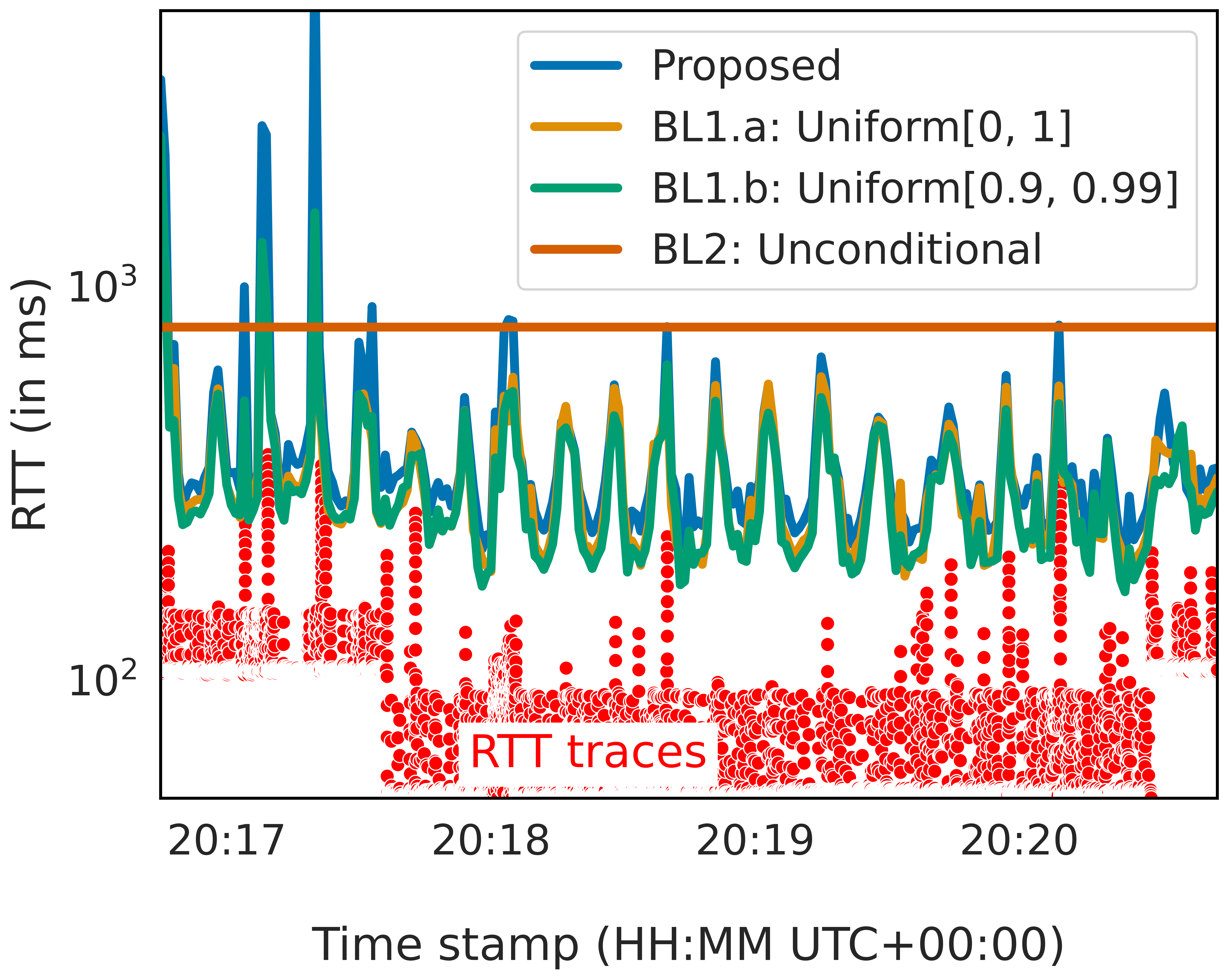}
    \caption{Snapshot of quantile forecasting showing the estimated $0.99$-quantile obtained with proposed estimator and baselines.}
    \label{fig:example-forecasting}
    \vspace{-1em}
\end{figure}

To illustrate quantile forecasting, consider the same time interval as illustrated in \Fig{example} (a time interval from the test set) and the goal of estimating the $0.99$-quantile of the \ac{RTT}. 
\Fig{example-forecasting} shows the $0.99$-quantiles obtained with our proposed high-quantile estimator (``Proposed'') as well as baselines (``BL''). 
Being agnostic to distributional shifts, the unconditional baseline often overestimates the $0.99$-quantile, while conditional quantile estimators (``Proposed'' and baselines 1.a and 1.b) adapt to distributional shifts caused by inter-beam and inter-satellite handovers. %, resulting in more flexible and desirable quantile estimates over time. 
In the following sections, we contrast these estimators based on their accuracy and performance. 

\subsection{Accuracy} \label{sec:accuracy}

In the absence of true conditional quantile values, we define \emph{accuracy} as the rate in which the predicted quantile $z_t$ is greater than or equal to the highest \ac{RTT} value observed during the horizon window, i.e., $\mathtt{accuracy} = \frac{1}{T}\sum_t^T \mathds{1}(z_t \ge y_t)$.
% \begin{equation}
%     \mathtt{accuracy} = \frac{1}{T}\sum_t^T \mathds{1}(z_t \ge y_t).
% \end{equation}
Naturally, for a reliability level of $\alpha = 0.99$, the accuracy should be  $0.99$, meaning that predictions for the $0.99$-quantile should not fail more than $1\%$ of the time. 
In other words, accuracy denotes a long-term measure of reliability, and an estimator performs as intended if its accuracy equals the reliability requirement.   

% As a reference, we compare the performance of the proposed method to the approach proposed in \cite{tagasovska2019single}: neural networks are trained with the expected pinball loss function with an uniformly distributed $\mathcal{A}$. 
% We consider two variations, $\mathcal{A} \sim \text{Uniform}[0, 1]$ and $\mathcal{A} \sim \text{Uniform}[0.9, 0.99]$. 

\begin{figure}[t]
    \centering
    \includegraphics[width=0.8\linewidth]{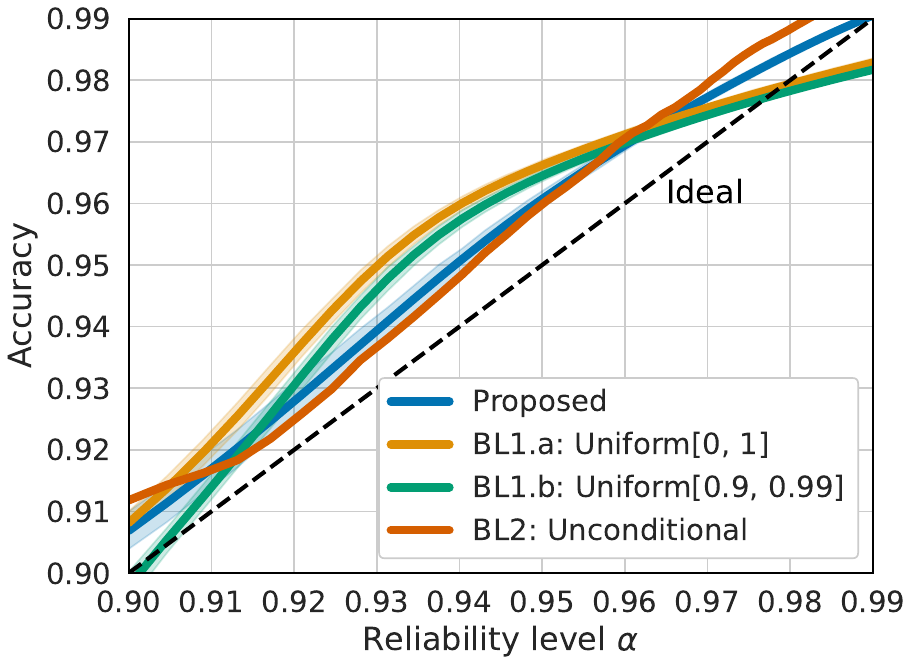}
    \caption{Accuracy as a function of the reliability requirement. 
    Solid lines show the average performance over independently trained estimators; shades, the $95\%$ confidence interval.}
    \label{fig:accuracy}
    \vspace{-1em}
\end{figure}

\Fig{accuracy} shows the accuracy of the proposed estimator and baselines for reliability requirements in the range of interest. 
The black dashed line is plotted as a reference and refers to the ideal case of $\mathtt{accuracy} = \alpha$. 
% Since the oracle ``global'' baseline corresponds to the $\alpha$-quantile computed over the entire testing set, its accuracy is very close to ideal, as expected.  
% Note that any major deviation from the ideal case is undesirable. 
If accuracy is less than $\alpha$ (i.e., curves fall under ``Ideal''), the estimator is obviously unable to meet reliability requirements in the long-term, which can jeopardize safety in remote control applications. 
If accuracy is significantly greater than $\alpha$ (i.e., curves are significantly above ``Ideal''), the estimator is likely conservative and overestimates $\alpha$-quantiles. 
It is important to note that, even though overestimation can support meeting reliability goals, it is undesirable in practice because it can reduce efficiency by ultimately slowing down the operational speed of the remote-controlled vehicle, as per definition in \Eqs{conditional-quantile}. 

The proposed estimator outperforms baselines 1.a and 1.b by \1 performing closer to ideal for most of the reliability range of interest and \2 always meeting reliability requirements (note how baselines 1.a and 1.b fall under ``Ideal'' as $\alpha \rightarrow 1$ while ``Proposed'' does not). 
In the following section, we discuss another facet of the problem and how the proposed estimator outperforms the unconditional estimator in practice. 

\subsection{Remote control performance}

\begin{figure}[t]
    \centering
    \includegraphics[width=0.8\linewidth]{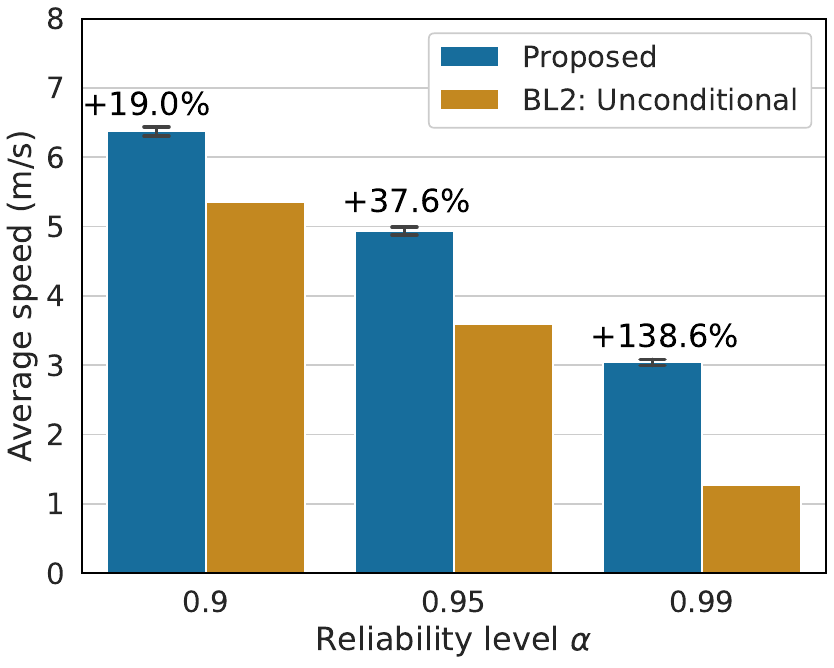}
    \caption{Average achieved speed at different reliability levels. Error bars indicate the $95\%$ confidence interval. }
    \label{fig:avg_speed}
    \vspace{-1em}
\end{figure}

Following the problem definition in \Sec{system-model}, the maximum speed a remote-controlled vehicle can safely operate is a function of the $\alpha$-quantile of the \ac{RTT} and a safety threshold $\epsilon$. 
% As a reference, let us consider the safety threshold of $\epsilon = 1~\si{\meter}$. 
\Fig{avg_speed} shows the average speed achieved with the proposed high-quantile estimator and the unconditional baseline estimator when the safety threshold is set to $\epsilon = 1~\si{\meter}$. 
Based on the observation from the previous section that baselines 1.a and 1.b are unable to meet long-term reliability requirements, they are omitted in this section, so we can focus on the relative gains obtained with the proposed estimator compared to the unconditional baseline. 

Our proposed high-quantile estimator yields significantly higher average speeds than the unconditional estimator. Specifically, the proposed estimator delivers average speeds that are 19\%, 37.6\%, and 138.6\% higher than the unconditional baseline for $\alpha \in \{0.9, 0.95, 0.99\}$, respectively. 
%Notably, the proposed estimator also outperforms the unconditional baseline even when unconditional quantiles are computed on the test set (label ``Test'').
The higher speeds obtained with the proposed estimator are due to its ability to forecast and adapt to time-varying changes in the statistics of the \ac{RTT}, such as those caused by inter-beam and inter-satellite handovers. 
In remote control, such higher speeds are desirable because they often map onto increased utilization of the remote-controlled vehicle. 
From \Fig{accuracy} and \Fig{avg_speed}, we observe that our proposed high-quantile estimator can support remote control by allowing higher speeds without sacrificing long-term reliability. 

\subsection{Training set size}

\begin{figure}
    \centering
    \includegraphics[width=0.8\linewidth]{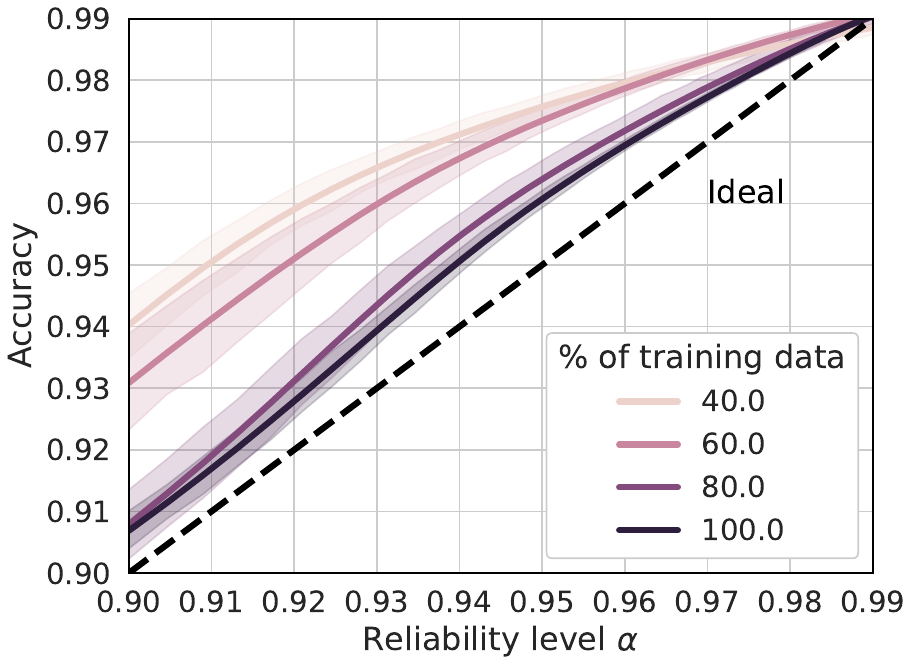}
    \caption{Accuracy as a function of the reliability requirement when estimators are trained with a fraction of the training set.}
    \label{fig:nsamples_training}
    \vspace{-1em}
\end{figure}

This section reports on the impact of the number of training samples on the proposed high-quantile estimator. 
Our training set contains $661,950$ $\{(\mathbf{x}_t, y_t)\}$ samples corresponding to measurements collected during February 2025. Combined, these samples correspond to roughly $7.6$ days of measurements. 
\Fig{nsamples_training} is \Fig{accuracy}'s counterpart and shows the accuracy of our proposed estimator when neural networks are trained only with a fraction of the samples in the training set.  
% As a reference, the dashed line shows the ideal case of $\mathtt{accuracy} = \alpha$. 

As expected, performance increases with more data—note how curves move towards ``Ideal'' as the number of training samples increases—with a small improvement from $80\%$ to $100\%$ (entire training set), which suggests that the proposed high-quantile estimator requires at least $\approx 6$ days of training data to perform satisfactorily. 

\section{Conclusion}

This paper studied how \ac{LEO} networks can support farm automation by formulating remote control as a quantile forecasting problem and proposing a high-quantile estimator to support remote control over the OneWeb \ac{LEO} network. 
We assessed the performance of the proposed high-quantile estimator using a real-world OneWeb dataset of \ac{RTT} and \ac{SINR} traces collected between a major agricultural hub and a major data center hub in the US. 
Our results show that the proposed estimator is attractive for remote control because it can meet long-term reliability goals while delivering operational speeds up to $138.6\%$ higher than baselines at the high-reliability levels of $99\%$, which is naturally beneficial for efficient and safe operation of remote-controlled farm vehicles.  

For future work, we plan to investigate if the proposed estimator can be augmented by tail-oriented frameworks such as extreme-value theory. 
We also plan to develop prototypes to further test our proposed remote control solution. %, e.g., using Iowa State University's ARA wireless living lab. % \cite{islam2025design}.  

\bibliographystyle{IEEEtran}
\bibliography{references}

@INPROCEEDINGS{dataset,
  author={Zhao, Jinwei and Perrin, Owen and Ahangarpour, Ali and Pan, Jianping},
  booktitle={2025 9th Network Traffic Measurement and Analysis Conference (TMA)}, 
  title={Measuring the {OneWeb} Satellite Network}, 
  year={2025},
  volume={},
  number={},
  pages={1-10},
  doi={10.23919/TMA66427.2025.11096999}}

@inproceedings{mohan2024multifaceted,
  title={A multifaceted look at {Starlink} performance},
  author={Mohan, Nitinder and Ferguson, Andrew E and Cech, Hendrik and Bose, Rohan and Renatin, Prakita Rayyan and Marina, Mahesh K and Ott, J{\"o}rg},
  booktitle={Proceedings of the ACM Web Conference 2024},
  pages={2723--2734},
  year={2024}
}

@inproceedings{pan2024measuring,
  title={Measuring the satellite links of a {LEO} network},
  author={Pan, Jianping and Zhao, Jinwei and Cai, Lin},
  booktitle={ICC 2024 - IEEE International Conference on Communications},
  pages={4439--4444},
  year={2024},
  organization={IEEE}
}

@article{hyndman1996sample,
  title={Sample quantiles in statistical packages},
  author={Hyndman, Rob J and Fan, Yanan},
  journal={The American Statistician},
  volume={50},
  number={4},
  pages={361--365},
  year={1996},
  publisher={Taylor \& Francis}
}

@misc{scipy_savgol_filter,
  author       = {SciPy},
  title        = {{Savitzky-Golay} filter},
  url          = {https://docs.scipy.org/doc/scipy/reference/generated/scipy.signal.savgol_filter.html},
  note         = {Accessed: 2026-04-23}
}

@inproceedings{dataset-ref2,
author = {Zhao, Jinwei and Pan, Jianping},
title = {{LENS}: A {LEO} Satellite Network Measurement Dataset},
year = {2024},
isbn = {9798400704123},
publisher = {Association for Computing Machinery},
address = {New York, NY, USA},
url = {https://doi.org/10.1145/3625468.3652170},
doi = {10.1145/3625468.3652170},
booktitle = {Proceedings of the 15th ACM Multimedia Systems Conference},
pages = {278–284},
numpages = {7},
location = {Bari, Italy},
series = {MMSys '24}
}

@article{tagasovska2019single,
  title={Single-model uncertainties for deep learning},
  author={Tagasovska, Natasa and Lopez-Paz, David},
  journal={Proceedings of the 33rd International Conference on Neural Information Processing Systems},
  volume={32},
  year={2019}
}

@article{santos2020agriculture,
  title={Agriculture 4.0--Agricultural robotics and automated equipment for sustainable crop production},
  author={Santos Valle, Santiago and Kienzle, Josef},
  year={2020},
  publisher={Food and Agriculture Organization of the United Nations}
}

@article{liu2025vivisecting,
  title={Vivisecting {Starlink} Throughput: Measurement and Prediction},
  author={Liu, Zikun and Reidys, Fan-Xue Gabriella and Tanveer, Sarah and Vasisht, Deepak},
  journal={Proceedings of the ACM on Networking},
  volume={3},
  number={CoNEXT4},
  pages={1--23},
  year={2025},
  publisher={ACM New York, NY, USA}
}

@inproceedings{zhao2024low,
  title={Low-latency live video streaming over a low-earth-orbit satellite network with dash},
  author={Zhao, Jinwei and Pan, Jianping},
  booktitle={Proceedings of the 15th ACM Multimedia Systems Conference},
  pages={109--120},
  year={2024}
}

@inproceedings{cao2023satcp,
  title={{SaTCP}: Link-layer informed {TCP} adaptation for highly dynamic {LEO} satellite networks},
  author={Cao, Xuyang and Zhang, Xinyu},
  booktitle={IEEE INFOCOM 2023-IEEE Conference on Computer Communications},
  pages={1--10},
  year={2023},
  organization={IEEE}
}

@article{expected-pinball,
    title={Expected Pinball Loss For Quantile Regression And Inverse {CDF} Estimation},
    author={Taman Narayan and Serena Lutong Wang and Kevin Robert Canini and Maya Gupta},
    journal={Transactions on Machine Learning Research},
    issn={2835-8856},
    year={2024},
    note={}
}

@misc{starlink-quantile-analysis,
      title={Statistical Characterization and Prediction of {E2E} Latency over {LEO} Satellite Networks}, 
      author={Andreas Casparsen and Jonas Ellegaard Jakobsen and Jimmy Jessen Nielsen and Petar Popovski and Israel Leyva Mayorga},
      year={2026},
      eprint={2601.08439},
      archivePrefix={arXiv},
      primaryClass={cs.NI},
      url={https://arxiv.org/abs/2601.08439}, 
}

@article{husseyn2025characterizing,
  title={Characterizing Latency Inflation in Mobile and Satellite Internet Connections},
  author={Husseyn, D and Saranya, DG and Babu, DK and Kishore, D and Kiruthikadevi, K},
  journal={Journal of Internet Services and Information Security},
  volume={15},
  pages={573--586},
  year={2025}
}

@INPROCEEDINGS{10228912,
  author={Ma, Sami and Chou, Yi Ching and Zhao, Haoyuan and Chen, Long and Ma, Xiaoqiang and Liu, Jiangchuan},
  booktitle={IEEE INFOCOM 2023-IEEE Conference on Computer Communications}, 
  title={Network Characteristics of {LEO} Satellite Constellations: A Starlink-Based Measurement from End Users}, 
  year={2023},
  volume={},
  number={},
  pages={1-10},
  doi={10.1109/INFOCOM53939.2023.10228912}}

@INPROCEEDINGS{Tian2025rttprediction,
  author={Tian, Jingyi and Yang, Wenjun and Cai, Lin},
  booktitle={ICC 2025-IEEE International Conference on Communications}, 
  title={Attention-Based Spatiotemporal Model for RTT Prediction in LEO Satellite Networks}, 
  year={2025},
  volume={},
  number={},
  pages={5059-5063},
  doi={10.1109/ICC52391.2025.11160911}}

@article{yaacoub2020key,
  title={A key {6G} challenge and opportunity—{Connecting} the base of the pyramid: A survey on rural connectivity},
  author={Yaacoub, Elias and Alouini, Mohamed-Slim},
  journal={Proceedings of the IEEE},
  volume={108},
  number={4},
  pages={533--582},
  year={2020},
  publisher={IEEE}
}

@inproceedings{lopez2023connecting,
  title={Connecting rural areas: an empirical assessment of {5G} terrestrial-LEO satellite multi-connectivity},
  author={L{\'o}pez, Melisa and Damsgaard, Sebastian Bro and Rodr{\'\i}guez, Ignacio and Mogensen, Preben},
  booktitle={2023 IEEE 97th Vehicular Technology Conference (VTC2023-Spring)},
  pages={1--5},
  year={2023},
  organization={IEEE}
}

\end{document}